\documentclass{article}
\usepackage{spconf,amsmath,amssymb,graphicx}
\usepackage{multirow}
\usepackage{bbm}
\usepackage{booktabs}
\usepackage{float}
\usepackage{hyperref}
\usepackage{multicol, blindtext}
\usepackage{enumitem}
\usepackage{comment}
\usepackage{graphicx}
 
\newcommand{\norm}[1]{\left \lVert #1 \right \rVert}
\newcommand{\argmin}{\mathop{\mathrm{arg\,min}}\limits}

\def\x{\ensuremath{\mathbf{x}}}

\def\z{\ensuremath{\mathbf{z}}}

\def\RR{\ensuremath{\mathbb{R}}}
\def\PP{\ensuremath{\mathbb{P}}}

\title{Unsupervised Speech Enhancement using Data-defined Priors}
\name{Dominik Klement$^1$, Matthew Maciejewski$^3$, Sanjeev Khudanpur$^{2,3}$, Jan Černocký$^1$, Lukáš Burget$^1$}
\address{$^1$Speech@FIT, Brno University of Technology, Czech Republic, \\ $^2$CLSP \& $^3$HLTCOE, Johns Hopkins University, USA}
\begin{document}
\ninept
\maketitle
\begin{abstract}
The majority of deep learning-based speech enhancement methods require paired clean-noisy speech data. Collecting such data at scale in real-world conditions is infeasible, which has led the community to rely on synthetically generated noisy speech. However, this introduces a gap between the training and testing phases. In this work, we propose a novel dual-branch encoder-decoder architecture for unsupervised speech enhancement that separates the input into clean speech and residual noise. Adversarial training is employed to impose priors on each branch, defined by unpaired datasets of clean speech and, optionally, noise. Experimental results show that our method achieves performance comparable to leading unsupervised speech enhancement approaches. Furthermore, we demonstrate the critical impact of clean speech data selection on enhancement performance. In particular, our findings reveal that performance may appear overly optimistic when in-domain clean speech data are used for prior definition---a practice adopted in previous unsupervised speech enhancement studies.
\end{abstract}
%
\section{Introduction}

Speech Enhancement (SE) seeks to recover the underlying clean speech given an observed noisy signal, corrupted by environmental noise, reverberation or transmission, by suppressing the noise and preserving as much speech information as possible. With the advent of deep learning, the community has focused on building SE systems based on deep neural networks (DNN)~\cite{xu2014regression, park2017fully,fu2019metricgan, segan}. These models are trained in a supervised fashion, requiring paired corpora of noisy and clean speech. However, collecting such data in reality is impractical at scale
. Hence, the community has been utilizing artificially simulated data\footnote{Clean speech summed with noise and reverberated}. This approach, however, introduces a mismatch between the training and test phase, as the model never sees the real noisy recordings during training.

To address it, unsupervised approaches have been recently studied. Fujimura et al.~\cite{fujimura_2021_noisy_target_training} introduced noisy-target training (NyTT) paradigm, consisting of adding one more source of noise to already noisy recording. Albeit allowing the use of real noisy data, the input is still artificial, not addressing the aforementioned mismatch between train and test time. Furthermore, MetricGAN-U~\cite{metricu_gan} was proposed as an extension to the previously introduced MetricGAN~\cite{fu2019metricgan} approach, where discriminator output is forced to correlate with a black-box non-intrusive metric DNSMOS~\cite{dnsmos}, which was recently extended in MOS-GAN~\cite{jiang2025_mos_gan}. On one hand, this family of methods does not require any clean speech data, on the other hand, the model performance is tied to the performance of the external models (e.g., DNSMOS) used to guide the training. To achieve better performance, one would need to re-train DNSMOS-like models, which is costly.

In contrast, unsupervised SE methods that require clean speech but do not require paired clean and noisy speech overcome these limitations. These methods usually use clean speech data as priors --- i.e., defining how the clean speech sounds like. Jiang et al. introduced unSE~\cite{jiang2023_unse} and unSE+~\cite{jiang2024_unse_plus}, a method that uses a discriminator to enforce the clean speech prior and computes reconstruction losses between the generated clean speech and the input noisy speech to ensure consistency. Despite the impressive results, optimizing a reconstruction loss between the generated clean speech and the input noisy speech requires careful tuning to prevent the model from reconstructing the input. 

In this work, we propose a novel dual-branch encoder-decoder architecture for unsupervised SE. Our architecture consists of two branches, which output clean speech and noise. We use adversarial training to impose branch-specific priors, and reconstruction losses between the noisy input and the combination of the two branches to ensure clean speech consistency. Therefore, there is less tension between the reconstruction and the adversarial training. Also, compared to the previous approaches, parallel branches allow general source separation of more than two distinct sources by adding more branches and utilizing source-specific data (e.g., piano music for instrument separation), which is left as a future work. 

We perform experiments to show how our method compares to the prior work and to show the importance of the discriminators. Furthermore, we show that initializing the model from a pre-trained neural audio codec (NAC)~\cite{defossez2023highfidelity} not only improves the convergence rate but also the performance of the enhancement model. Lastly, we show how important is the choice of the clean speech data and how one needs to be careful when selecting it in order to achieve the best performance. The codebase is released on GitHub \footnote{\href{https://github.com/BUTSpeechFIT/USE_DDP}{https://github.com/BUTSpeechFIT/USE\_DDP}}.

\vspace{-10pt}
\section{Proposed Method}
\begin{figure*}[t]
    \centering
    \includegraphics[width=.9\linewidth]{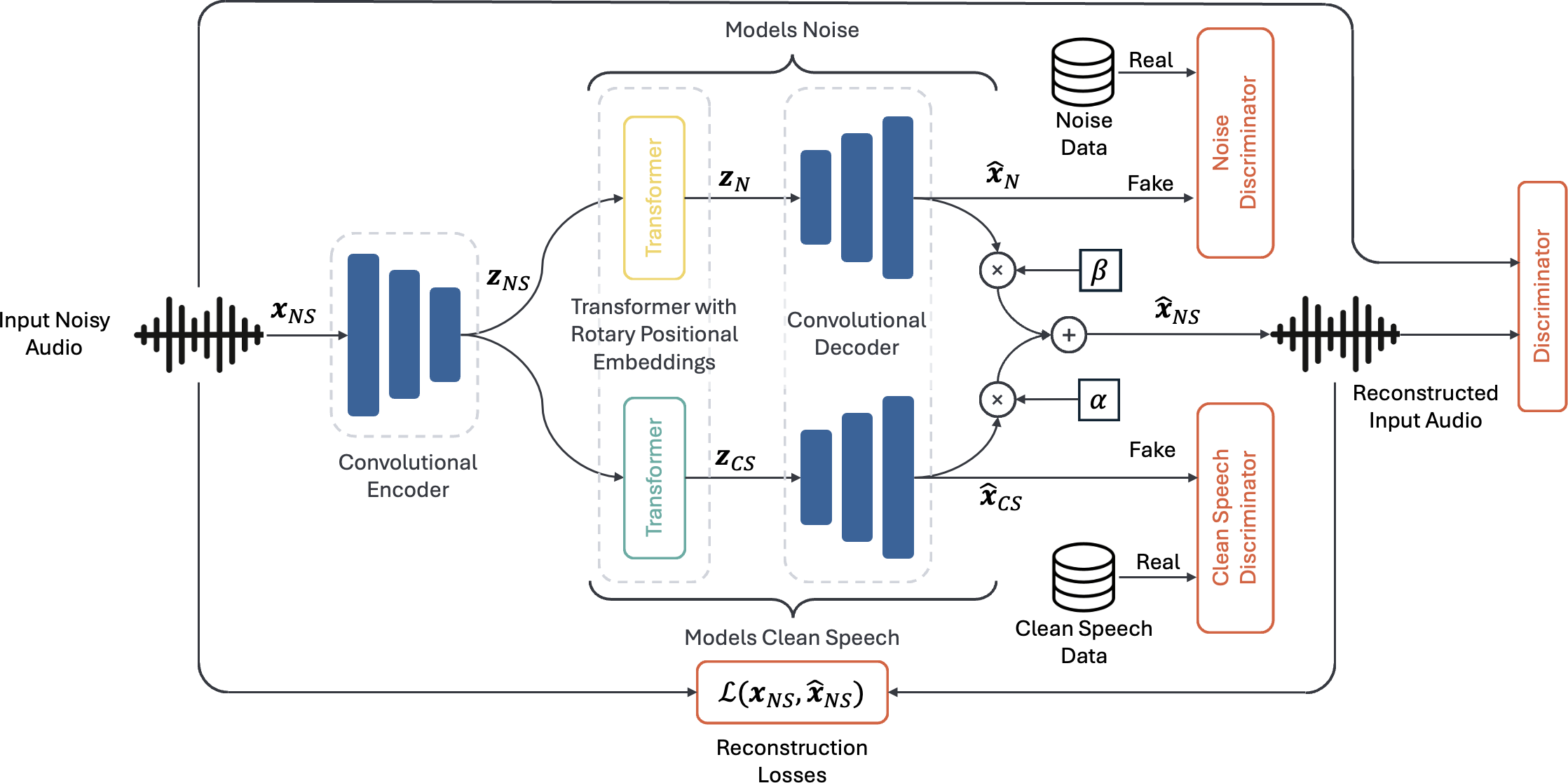}
    \vspace{-7pt}
    \caption{Diagram of the proposed dual-brach architecture.}
    \vspace{-5pt}
    \label{fig:model}
\end{figure*}

The proposed method is illustrated in Figure~\ref{fig:model}. It is an adversarially-trained model consisting of a generator and a set of discriminators. The generator is an encoder-decoder architecture that performs audio reconstruction task, similar to NAC. Internally, the proposed method consists of two branches that output clean speech and noise audio. The combination of the two audio signals is forced to reconstruct the input. To force the model to separate the input into two { \it distinct } source signals, we use adversarial training with source-specific discriminators to impose prior distributions on each branch. To train each discriminator, we use two independent corpora --- clean speech and noise, defining how branch-specific outputs should sound.

\subsection{Generator Architecture}
Let $\x_{NS} \in \RR^T$ be a noisy input audio of length $T$ samples. First, a convolutional encoder $E$ is applied to map the noisy input to a latent representation $\z_{NS} = E(\x_{NS}) \in \RR^{L \times M}$, where $L$ is the latent sequence length and $M$ is the latent space dimensionality. Subsequently, the latent representation is split into two parallel branches and transformer with rotary positional embeddings (RoFormer)\footnote{Positional information is important for model convergence.}~\cite{roformer} is applied, enabling each branch to capture distinct information. As a result, we obtain latent sequences $\z_N = \mathcal{R}_N(\z_{NS}), \z_{CS} = \mathcal{R}_{CS}(\z_{NS}) \in \RR^{L \times M}$, representing noise and clean latent representations respectively. Next, each branch is decoded back to the time domain using a single shared decoder $Dec$, resulting in $\hat{\x}_{N} = Dec(\z_N), \hat{\x}_{CS} = Dec(\z_{CS}) \in \RR^T$. Finally, the reconstructed noisy input is a combination of the two audio signals:

\vspace{-13pt}
\begin{gather}
    \hat{\x}_{NS} = \alpha^* \hat{\x}_{N} + \beta^* \hat{\x}_{CS}, \label{eq:noisy_comb} \\
    \alpha^*, \beta^* = \argmin_{\alpha, \beta} \norm{\x_{NS} - (\alpha \hat{\x}_{N} + \beta \hat{\x}_{CS})}_2^2 \label{eq:noisy_comb_alphabeta} 
\end{gather}
where $\alpha^*, \beta^* \in \RR$ are scalar weights accounting for errorneous amplitude estimation\footnote{We do not want to penalize the model for incorrectly estimating the amplitude, as it does not change the audio itself.}. The optimal values are obtained from a closed-from solution to the convex optimization problem defined above. In the following text, we will use $\hat{\x}_{CS} = G_{CS}(\x_{NS}), \hat{\x}_{N} = G_{N}(\x_{NS}), \hat{\x}_{NS} = G_{NS}(\x_{NS}) \in \RR^T$ to denote the generated clean speech, noise, and noisy input signal, respectively.

\vspace{-5pt}
\subsection{Adversarial Training}

Let $\PP_{CS}$ and $\PP_{N}$ be the prior distributions of real clean speech and noise, respectively, and $\PP_{NS}$ be the distribution of the noisy input. Our goal is to train the model such that $p(\hat{\x}_{CS}) \approx \PP_{CS}$ and $p(\hat{\x}_{N}) \approx \PP_{N}$. It has been shown by Goodfellow et al.~\cite{goodfellow2014generative} that if the generator and the discriminator have enough capacity, the distribution of the generated samples converges to the distribution of the real data. In our setup, we use 
three discriminators $D_{CS}, D_{N}, D_{NS}$, for clean speech, noise, and noisy speech, respectively. Each discriminator is an ensamble of multiple sub-discriminators (described in Section~\ref{sec:exp_setup_model}); however, we omit the summation of losses across the particular sub-discriminators for clarity. While the first two discriminators impose priors on the distributions of branch-wise output signals, the noisy speech one has the same role as in NACs --- i.e., improve fidelity of the reconstructed audio. Instead of the vanilla GAN setup~\cite{goodfellow2014generative}, we use LS-GAN loss~\cite{2017_mao_lsgan}. The discriminators are trained using the following loss function:
\vspace{-5pt}
\begin{equation}
    \begin{aligned}
        \mathcal{L}_{D_{CS}} = \mathbb{E}_{\substack{\x_{CS} \sim \PP_{CS}, \x_{NS} \sim \PP_{NS}}} \big[
        &D_{CS}(G_{CS}(\x_{NS}))^2 \\
        &+ (1 - D_{CS}(\x_{CS}))^2 \big].
    \end{aligned}
    \vspace{-2pt}
    \label{eq:lsgan_cs}
\end{equation}
$\mathcal{L}_{D_N}$ is defined analogously using $D_N$ and $\PP_N$. For input reconstruction, we train the noisy speech discriminator $D_{NS}$ using the same loss function but only one data distribution:
\vspace{-5pt}
\begin{equation}
    \vspace{-5pt}
    \begin{aligned}
        \mathcal{L}_{D_{NS}} = \mathbb{E}_{\substack{\x_{NS} \sim \PP_{NS}}} \big[D_{NS}(\hat{\x}_{NS})^2 + (1 - D_{NS}(\x_{Ns}))^2 \big].
    \end{aligned}
    \label{eq:lsgan_ns}
\end{equation}

\subsection{Generator Losses}
We follow the general NAC training~\cite{defossez2023highfidelity, kumar2023highfidelityaudiocompressionimproved} to train the generator for audio reconstruction task. We use multi-scale mel spectrogram loss~\cite{defossez2023highfidelity} and SI-SDR loss~\cite{sisdr}, and denote their combination as $\mathcal{L}_{rec}(\x_{NS})$ in the following text. In addition to reconstruction losses, we use adversarial loss and feature matching loss~\cite{kumar2023highfidelityaudiocompressionimproved} to increase the fidelity of the generated noisy speech audio:
\vspace{-7pt}
\begin{equation}
    \hspace{-1pt}
    \label{eq:feat_matching_loss}
    \mathcal{L}_{feat}(\x_{NS}, \hat{\x}_{NS}) = \sum_{k=1}^{K_D} \sum_{l=1}^{M_k-1} \left\lVert F_{k, l}(\x_{NS})-F_{k,l}(\hat{\x}_{NS}) \right\rVert_1, 
    \vspace{-2pt}
\end{equation}
where $K_D$ is the number of sub-discriminators, $M_k$ is the number of convolutional layers in $k$-th sub-discriminator, and $F_{k,l}$ denotes a convolutional feature map. 

As we do not have reference for the corresponding clean speech and noise, we only use adversarial losses to learn the source separation, defined as:
\vspace{-5pt}
\begin{equation}
    \mathcal{L}_{G_{CS}}(\x_{NS}) = \mathbb{E}_{\substack{\x_{NS} \sim \PP_{NS}}} [(1-D_{CS}(G_{CS}(\x_{NS})))^2],
    \vspace{-5pt}
\end{equation}
where $\mathcal{L}_{G_N}$ and $\mathcal{L}_{G_{NS}}$ are defined analogously.

To stabilize the training and avoid trivial mode collapse\footnote{A scenario where clean speech becomes silence and the entire noisy input is reconstructed through the noise branch.}, we add a regularization loss that maximizes the energy of the generated clean speech audio signal, defined as:
\begin{equation}
    \mathcal{L}_{emax}(\x_{NS}) = - \log \left( \frac{1}{T} \sum_{i=1}^T G_{CS}(\x_{NS})[i]^2\right),
\end{equation}
where $G_{CS}(\x_{NS})[i]$ denotes $i$-th clean speech signal sample.

The overall generator loss function is a weighted sum of the noisy reconstruction and adversarial losses, and branch-specific adversarial losses, defined as:
\begin{equation}
    \begin{aligned}
        \mathcal{L}_G &= \lambda_{gcs}\mathcal{L}_{G_{CS}} + \lambda_{gn} \mathcal{L}_{G_N} + \lambda_{gns}\mathcal{L}_{G_{NS}} + \lambda_{feat} \mathcal{L}_{feat} \\
        &+ \lambda_{rec} \mathcal{L}_{rec} + \lambda_{emax} \mathcal{L}_{emax}.
    \end{aligned}
\end{equation}



\subsection{Why does it work?}
Adversarial training with branch-specific discriminators constrains each branch to follow the distribution defined by the corresponding data, i.e., 
$p(\hat{\mathbf{x}}_{CS}) \approx \mathbb{P}_{CS}$ and $p(\hat{\mathbf{x}}_{N}) \approx \mathbb{P}_{N}$. 
Simultaneously, the combination of the two outputs must reconstruct the original noisy input. 
This dual constraint prevents the clean speech branch from generating arbitrary clean speech solely to satisfy the adversarial objective, 
as such signal would increase the reconstruction loss. 
Consequently, the clean branch is compelled to generate an enhanced version of the noisy input---one that both aligns with the clean speech prior 
and contributes to accurate input reconstruction.

\vspace{-7pt}
\section{Experimental Setup}
\vspace{-3pt}

\subsection{Model}
\label{sec:exp_setup_model}
We built the model architecture on top of the Descript Audio Codec (DAC)~\cite{kumar2023highfidelityaudiocompressionimproved}. The encoder contains four residual blocks with downsampling factors $[2, 4, 5, 8]$, resulting in 320 downsampling ratio (50Hz for 16kHz input) and 1024-dimensional embeddings. The decoder mirrors the encoder. Furthermore, each branch contains a transformer with 8 layers, 8 attention heads, and feed-forward dimensionality of 1536. If not specified otherwise, we initialize the encoder and decoder weights from a pre-trained DAC.

We use three discriminator ensembles. The noisy input reconstruction one, $D_{NS}$, exactly follows the DAC setup~\cite{kumar2023highfidelityaudiocompressionimproved}---multi-period (MPD) with periods $[2,3,5,7,11]$ and multi-band multi-scale STFT (MS-STFT) discriminator with 5 bands, STFT window sizes $[2048, 1024, 512]$, hop length $\frac{1}{4}$ of the window size, and 32 convolution filters . For the branch-specific discriminators that define speech and noise priors, we use a single-band MS-STFT discriminator, as we observed that using MPD did not improve the performance, with STFT window sizes $[2048, 1024, 512, 256, 128]$ and 128 convolution filters.
\vspace{-5pt}
\subsection{Training}
We optimize our models using AdamW~\cite{Loshchilov2017DecoupledWD} with weight decay of $0.02$.. We train all models for up to 50k iterations, each consisting of first updating the discriminator and then the generator. The learning rate (LR) is linearly warmed up for 5k iterations, achieving peak LR 2e-4, and then decayed according to the cosine scheduler throughout the rest of the training. We use gradient clipping of 1.0 and mixed precision training in bfloat16. The best-performing model is selected based on the validation set.

\subsection{Data}
We focus on 16kHz sampling rate and all the datasets are resampled before processing. For the comparison with prior work and clean speech prior selection experiments, we use VCTK+Demand (VD)~\cite{vctk_demand} subset containing 28 speakers for training (out of which two speakers are reserved for validation), and the standard 2 speakers test set. For the ablations, in order to not base our architectural decisions on a small dataset, we utilize URGENT challenge Task1~\cite{urgent_challenge} data setup consisting of multiple speech and noise corpora, where we randomly select 1000 utterances for validation. The test set consists of 2368 recordings simulated from per-corpora test sets\footnote{We only use reverberation and noise, compared to the URGENT challenge where other types of distortions were used.}. To show the performance on real noisy data and to further explore the clean speech prior data selection, we use CHIME-3~\cite{barker2015_chime3} dataset, comprised of pre-defined train, validation, and test splits.

During training, we always use the noise corpora defined by the URGENT challenge. For clean speech, we use the VD clean train set to be comparable with prior work, even though it does not resemble the conditions the model is going to be exposed to during real noisy data training (i.e., prior mismatch between the clean speech data and the underlying clean speech the model is trained to estimate). To address this problem, we explore the clean speech prior data selection, where we use the URGENT clean speech data setup without VCTK to simulate out-of-domain clean speech data.

\vspace{-5pt}
\subsection{Metrics}
To evaluate the model performance and compare to prior works, we follow the setups defined in~\cite{metricu_gan,jiang2023_unse} and use DNSMOS~\cite{dnsmos} to judge the quality of the enhanced speech (cleanliness), PESQ~\cite{pesq} to evaluate perceptual speech quality, and CSIG, CBAK, COVL (based on PESQ)~\cite{hu2008_eval} to obtain a predicted MOS for: speech, background noise, and overall speech quality. For real noisy data evaluations and clean speech prior data selection experiments, we use DNSMOS~\cite{dnsmos} and UTMOS~\cite{utmos}. 

\vspace{-10pt}
\section{Results}

\subsection{Comparison to Prior Work}
We compare our method to previously published methods using the VD dataset and use the VCTK clean speech for adversarial training to be comparable to the prior approaches. We acknowledge that this setup leads to over-optimistic results; therefore, we explore on out-of-domain clean speech data later in Section~\ref{sec:prior_selection}.

\vspace{-5pt}
\begin{table}[hbt]
    \centering
    \caption{Comparison of our unsupervised models with prior work.}
    \label{tab:prior_unsup_work_comparison}
    \addtolength{\tabcolsep}{-4.3pt}
    \begin{tabular}{@{}lccccc@{}}
        \toprule  
 Model & DNSMOS$\uparrow$ & PESQ$\uparrow$ & CSIG$\uparrow$ & CBAK$\uparrow$ & COVL$\uparrow$ \\ 
        \midrule
        Input Speech & 2.54 & 1.97 & 3.35 & 2.44 & 2.63 \\
                \midrule
         NyTT~\cite{fujimura2023_nytt} & - & 2.30 & 3.19 & 3.01 & 2.72 \\
         MGAN-U (half)~\cite{metricu_gan} & 2.89 & 2.45 & 3.47 & 2.63 & 2.91 \\
         MGAN-U (full)~\cite{metricu_gan} & \textbf{3.15} & 2.13 & 3.22 & 2.42 & 2.63 \\
         MOS-GAN~\cite{jiang2025_mos_gan} & 2.91 & 2.40 & - & - & - \\
         unSE~\cite{jiang2023_unse} & 2.92 & 2.45 & \textbf{3.69} & 3.05 & \textbf{3.05} \\
         unSE+~\cite{jiang2024_unse_plus} & 2.94 & \textbf{2.48} & 3.31 & \textbf{3.07} & 2.85 \\
                \midrule
         Ours & 3.03 & 2.47 & 3.54 & 2.35 & 2.99 \\ 
        \bottomrule
    \end{tabular}
\end{table}

Table~\ref{tab:prior_unsup_work_comparison} shows that all methods outperform the input noisy speech, suggesting they are capable of enhancing the noisy speech. Our method outperforms all but MetricGAN-U (full) in terms of DNSMOS; however, the full model used DNSMOS during training, which strenghtens its DNSMOS performance. Furthermore, our method achieves the second best results in PESQ, CSIG and COVL, suggesting the quality of the enhanced speech is similar or better than unSE---a method our model is the most comparable to. Lastly, despite the good enhancement performance, the audio sometimes contains some noise or noisy artifact, which is reflected in lower CBAK. Overall, the fairest baselines are unSE and unSE+, which our method performs on par with.

\vspace{-5pt}
\subsection{Initialization}
Table~\ref{tab:unsup_init_metrics} compares the performance of our method when the encoder and the decoder are initialized from a pre-trained NAC and when trained from scratch. It can be seen that initializing from a pre-trained NAC improves the performance as the encoder already produces meaningful general audio latent representations that branch-wise transformers learn early to separate. Also, the model converges faster, requiring around 3x less training iterations.
\begin{table}[hbt]
    \centering
    \caption{Comparison of encoder and decoder initialization methods.}
    \label{tab:unsup_init_metrics}
    \addtolength{\tabcolsep}{-4.5pt}
    \begin{tabular}{@{}lccccc@{}}
        \toprule
        Init Method & DNSMOS~$\uparrow$ & PESQ~$\uparrow$ & CSIG~$\uparrow$ & CBAK~$\uparrow$ & COVL~$\uparrow$ \\
        \midrule
        From scratch & 2.28 & 1.41 & 2.08 & 1.44 & 1.66 \\
        Pretrained (NAC) & \textbf{2.60} & \textbf{1.45} & \textbf{2.13} & \textbf{1.79} & \textbf{1.70} \\
        \bottomrule
    \end{tabular}
\end{table}

\vspace{-10pt}
\subsection{Discriminator Ablations}
We investigate whether all three discriminators are required for optimal performance. The results are reported in Table~\ref{tab:unsup_disc_comparison}. When all discriminators are employed, the model achieves the highest scores in DNSMOS, CSIG, and CBAK. However, performance in PESQ and COVL is slightly lower compared to the model without the noise discriminator. This indicates that using all discriminators generally yields cleaner audio, but in some cases leads to excessive suppression. The underlying reason is that the noise discriminator enables the system to estimate noise more accurately, particularly in non-speech segments. Consequently, the clean speech branch is encouraged, through the reconstruction loss, to attenuate noise more aggressively in these regions, which occasionally results in partial suppression of the speech signal.

Furthermore, removing the noisy speech discriminator slightly reduces the perceptual quality of the enhanced speech. This discriminator plays an important role in ensuring that the reconstructed input audio maintains naturalness and fidelity, which impacts the clean speech branch as well. Without it, the enhanced speech often lacks high-frequency details, making the audio sound less natural and degraded.

\begin{table}[hbt]
    \centering
    \caption{Comparison of discriminators employed during training. The first row shows the model with all discriminators, the second one without the noise discriminator, and the last one only with the clean speech discriminator.}
    \label{tab:unsup_disc_comparison}
    \addtolength{\tabcolsep}{-3.7pt}

    \begin{tabular}{lccccc}
        \toprule
        Discriminators & DNSMOS~$\uparrow$ & PESQ~$\uparrow$ & CSIG~$\uparrow$ & CBAK~$\uparrow$ & COVL~$\uparrow$ \\
        \midrule
        all & \textbf{2.60} & 1.45 & \textbf{2.13} & \textbf{1.79} & 1.70 \\
        -noise & 2.46 & \textbf{1.48} & \textbf{2.13} & 1.73 & \textbf{1.72} \\
        -noise \& noisy & 2.52 & 1.43 & 2.05 & 1.72 & 1.66 \\
        \bottomrule
    \end{tabular}
\end{table}

\vspace{-8pt}
\subsection{Prior Selection}
\label{sec:prior_selection}

The clean speech data pool dictates how clean speech sounds. In this section, we explore how the data choice affects the model performance and how over-promising the performance may seem if in-domain clean speech data are used. In our experiments, we refer to data that either match the underlying clean speech data the model is trained to estimate (in the case of simulated data) or significantly correlate with the noisy speech data (e.g., close talk microphone) as in-domain. 

\vspace{-4pt}
\begin{table}[h]
    \centering
    \caption{Comparison of indomain and out-of-domain clean speech data when trained on VCTK+Demand dataset.}
    \label{tab:vctk_prior_comparison}

    \addtolength{\tabcolsep}{-3.7pt}
    \begin{tabular}{lccccc}
    \toprule
    CS Prior & DNSMOS~$\uparrow$ & PESQ~$\uparrow$ & CSIG~$\uparrow$ & CBAK~$\uparrow$ & COVL~$\uparrow$ \\
    \midrule
    in-domain & \textbf{3.03} & \textbf{2.47} & \textbf{3.54} & \textbf{2.35} & \textbf{2.99} \\ 
    out-of-domain & 2.86 & 2.04 & 2.60 & 2.06 & 2.27 \\
    \bottomrule
    \end{tabular}
\end{table}

Table~\ref{tab:vctk_prior_comparison} shows the clean speech prior comparison on VD. It can be seen that using an in-domain clean speech prior VCTK clean train set significantly outperforms the out-of-domain prior constructed from the URGENT clean speech pool not containing VCTK. This is mainly due to the distribution mismatch, allowing noise to leak into the clean speech branch. Since we do not have the ground-truth clean speech available when training on real noisy data, we consider using the in-domain prior for data-driven SE as unfair, as it significantly biases the model performance compared to the training scenario where in-domain clean speech data are unavailable.

\vspace{-5pt}
\begin{table}[h]
    \centering
    \caption{Comparison of in-domain and out-of-domain clean speech data when trained on CHiME-3 dataset.}
    \label{tab:chime3_prior_comparison}

    \addtolength{\tabcolsep}{-3.7pt}
    \begin{tabular}{lcccc}
    \toprule
    \multirow{2}{*}{CS Prior} & \multicolumn{2}{c}{Dev} & \multicolumn{2}{c}{Test} \\
     & DNSMOS~$\uparrow$ & UTMOS~$\uparrow$ & DNSMOS~$\uparrow$ & UTMOS~$\uparrow$ \\
    \midrule
    in-domain & 2.13 & 1.79 & 2.01 & 1.50 \\ 
    out-of-domain & \textbf{2.98} & \textbf{2.43} & \textbf{2.79} & \textbf{1.85} \\
    \bottomrule
    \end{tabular}
\end{table}



\begin{figure}[h!]
    \centering
    \begin{tabular}{@{}c@{}c@{}c@{}c@{}}
        \includegraphics[width=0.32\linewidth]{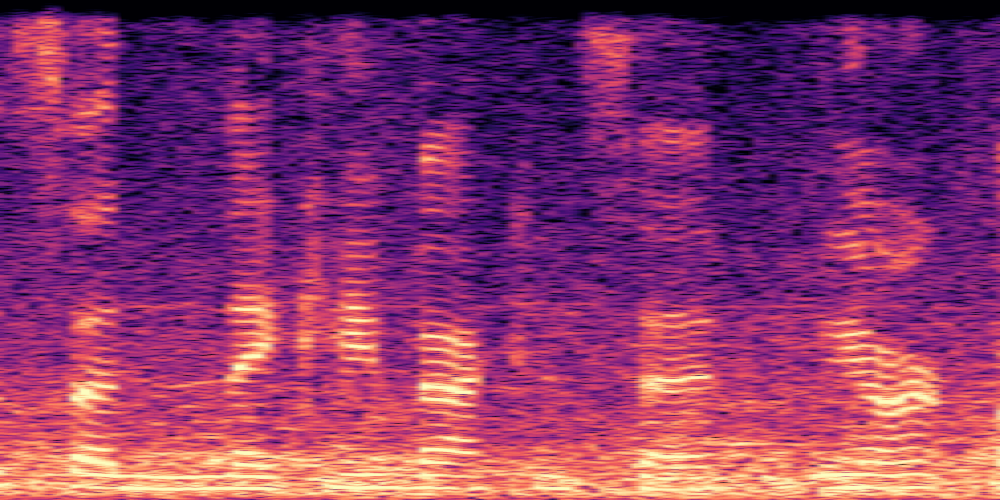} &
        \hspace{0.01\linewidth}
        \includegraphics[width=0.32\linewidth]{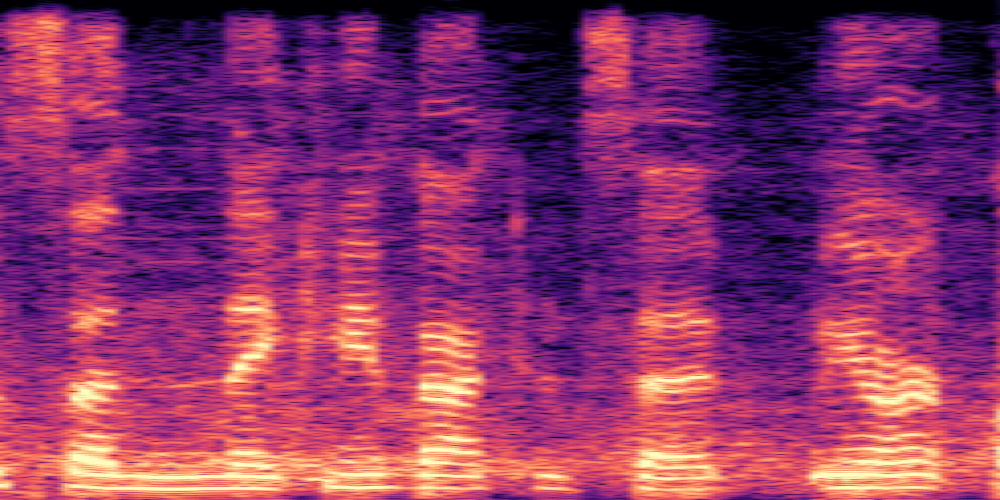} &
        \hspace{0.01\linewidth}
        \includegraphics[width=0.32\linewidth]{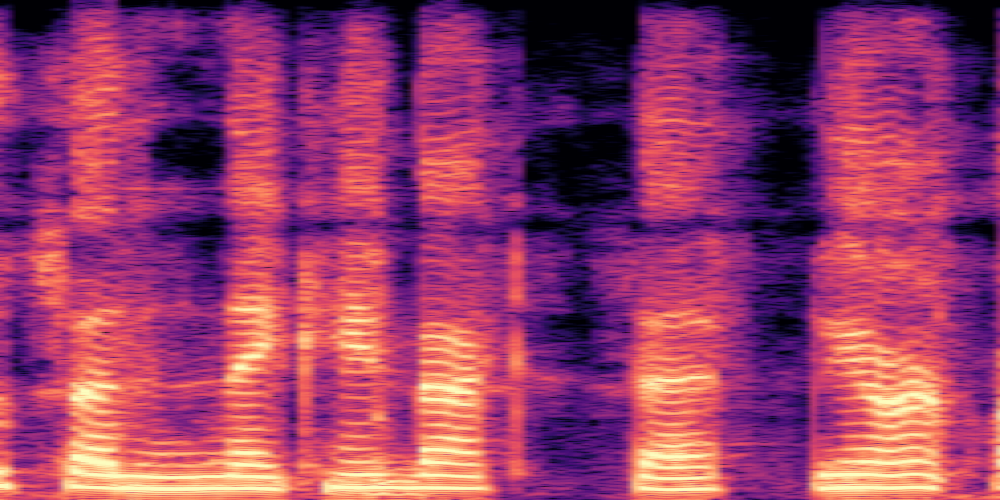}
    \end{tabular}
    \caption{CHiME-3 noisy, enhanced using in-domain and out-of-domain clean speech priors, respectively.}
    \label{fig:chime3_prior_comparison}
\end{figure}

Furthermore, Table~\ref{tab:chime3_prior_comparison} performs similar comparison on CHiME-3 real data, where we use channel 0 (close talk microphone) as the in-domain clean speech and the URGENT clean speech as the out-of-domain one. We can see the opposite effect compared to Table~\ref{tab:vctk_prior_comparison}, as the in-domain data are recorded in noisy environment using a close-talk microphone, resulting in a noisy ``clean speech`` prior. Therefore, the model leaks more noise to the signal compared to the out-of-domain prior, which contains only clean speech (see Figure~\ref{fig:chime3_prior_comparison}). However, due to the prior mismatch, the model tends to oversupress the noise at the expense of intelligibiliy.

\vspace{-14pt}
\section{Conclusion}
This work introduced a novel dual-branch encoder-decoder architecture for unsupervised speech enhancement, able to be trained on real noisy data. Compared to the MOS-based approaches, our approach relies solely on data and does not require any external model to guide the training. Our experiments show that the method is comparable to the top-performing prior unsupervised SE works. Furthermore, we pointed out the importance of the clean speech data selection and how it might mislead the performance evaluation. In the future, we plan to extend the method to unsupervised multi-source separation and investigate discriminator architecture to strengthen the priors and improve the robustness.

\vfill\pagebreak

\section{Acknowledgments}
This work was partially supported by NSF CCRI Grant No 2120435 and corporate gifts to Johns Hopkins University for JSALT 2025.  Computing on IT4I supercomputer was supported by MoE through the e-INFRA CZ (ID:90254).

\bibliographystyle{IEEEbib}
\bibliography{refs}

\end{document}